\documentclass[11pt]{article}

\usepackage[]{coling}

\usepackage{times}
\usepackage{latexsym}
\usepackage{xcolor}
\usepackage{colortbl}
\usepackage{float}
\usepackage{microtype}
\usepackage{graphicx}
\usepackage{inconsolata}
\usepackage{rotating}
\usepackage{adjustbox}
\usepackage{booktabs}%
\usepackage{algorithm}%
\usepackage{algorithmicx}%
\usepackage{algpseudocode}%
\usepackage{listings}%
\usepackage{subcaption}
\usepackage[T1]{fontenc}
\usepackage{graphicx}%
\usepackage{multirow}%
\usepackage{amsmath,amssymb,amsfonts}%
\usepackage{amsthm}%
\usepackage{mathrsfs}%
\usepackage[title]{appendix}%
\usepackage{xcolor}%
\usepackage{textcomp}%
\usepackage{manyfoot}%
\usepackage{booktabs}%
\usepackage{algorithm}%
\usepackage{algorithmicx}%
\usepackage{algpseudocode}%
\usepackage{listings}%
\usepackage{subcaption}
\usepackage{times}
\usepackage{latexsym}
\usepackage{dialogue}
\usepackage{xcolor}
\usepackage{colortbl}
\usepackage[utf8]{inputenc}
\usepackage{array} 
\usepackage{booktabs} 
\usepackage{microtype}
\usepackage{graphicx}
\usepackage{inconsolata}
\usepackage{rotating}
\usepackage{adjustbox}

\usepackage{float}

\definecolor{mygray}{gray}{0.9}
\definecolor{cellcolor}{rgb}{0,0,0} 
\definecolor{verdechiarissimissimo}{rgb}{0.95,1,0.95}
\definecolor{verdechiarissimo}{rgb}{0.88,1,0.88} 
\definecolor{verdechiaro}{rgb}{0.6,.9,0.6}
\definecolor{verdescuro}{rgb}{0.3,.7,0.3}
\definecolor{verdescurissimo}{rgb}{0.1,.5,0.1}

\usepackage[utf8]{inputenc}
\usepackage{amsmath}
\usepackage{microtype}
\usepackage{hyperref}

\usepackage{inconsolata}

\usepackage{graphicx}

\title{The Emotional Spectrum of LLMs: Leveraging Empathy and Emotion-Based Markers for Mental Health Support}

\newcommand*\samethanks[1][\value{footnote}]{\footnotemark[#1]}

\author{Alessandro De Grandi \thanks{The first two authors contributed equally to this work.} \\  Università della Svizzera italiana \\ \texttt{alessandro.de.grandi@usi.ch}
        \And 
        Federico Ravenda \samethanks \\  Università della Svizzera italiana \\ \texttt{federico.ravenda@usi.ch}
         \AND
        Andrea Raballo \\ Università della Svizzera italiana \\ \texttt{andrea.raballo@usi.ch}
                \And 
        Fabio Crestani \\  Università della Svizzera italiana \\ \texttt{fabio.crestani@usi.ch}}

\begin{document}
\maketitle

\begin{abstract}

The increasing demand for mental health services has highlighted the need for innovative solutions, particularly in the realm of psychological conversational AI, where the availability of sensitive data is scarce.
In this work, we explored the development of a system tailored for mental health support with a novel approach to psychological assessment based on explainable emotional profiles in combination with empathetic conversational models, offering a promising tool for augmenting traditional care, particularly where immediate expertise is unavailable.
Our work can be divided into two main parts, intrinsecaly connected to each other. First, we present RACLETTE, a conversational system that demonstrates superior emotional accuracy compared to considered benchmarks in both understanding users' emotional states and generating empathetic responses during conversations, while progressively building an emotional profile of the user through their interactions. Second, we show how the emotional  profiles of a user can be used as interpretable markers for mental health assessment. These profiles can be compared with characteristic emotional patterns associated with different mental disorders, providing a novel approach to preliminary screening and support.
\end{abstract}

\section{Introduction}

Empathetic chatbots represent a significant evolution in the field of conversational AI, designed not just to understand commands or queries, but to perceive and interpret the emotional states of their users. 
These advanced agents leverage Natural Language Processing (NLP) approaches to analyze text for emotional content, enabling them to engage in interactions that feel more human-like. 
\begin{figure}[ht]
    \centering
    \includegraphics[width=1.\linewidth]{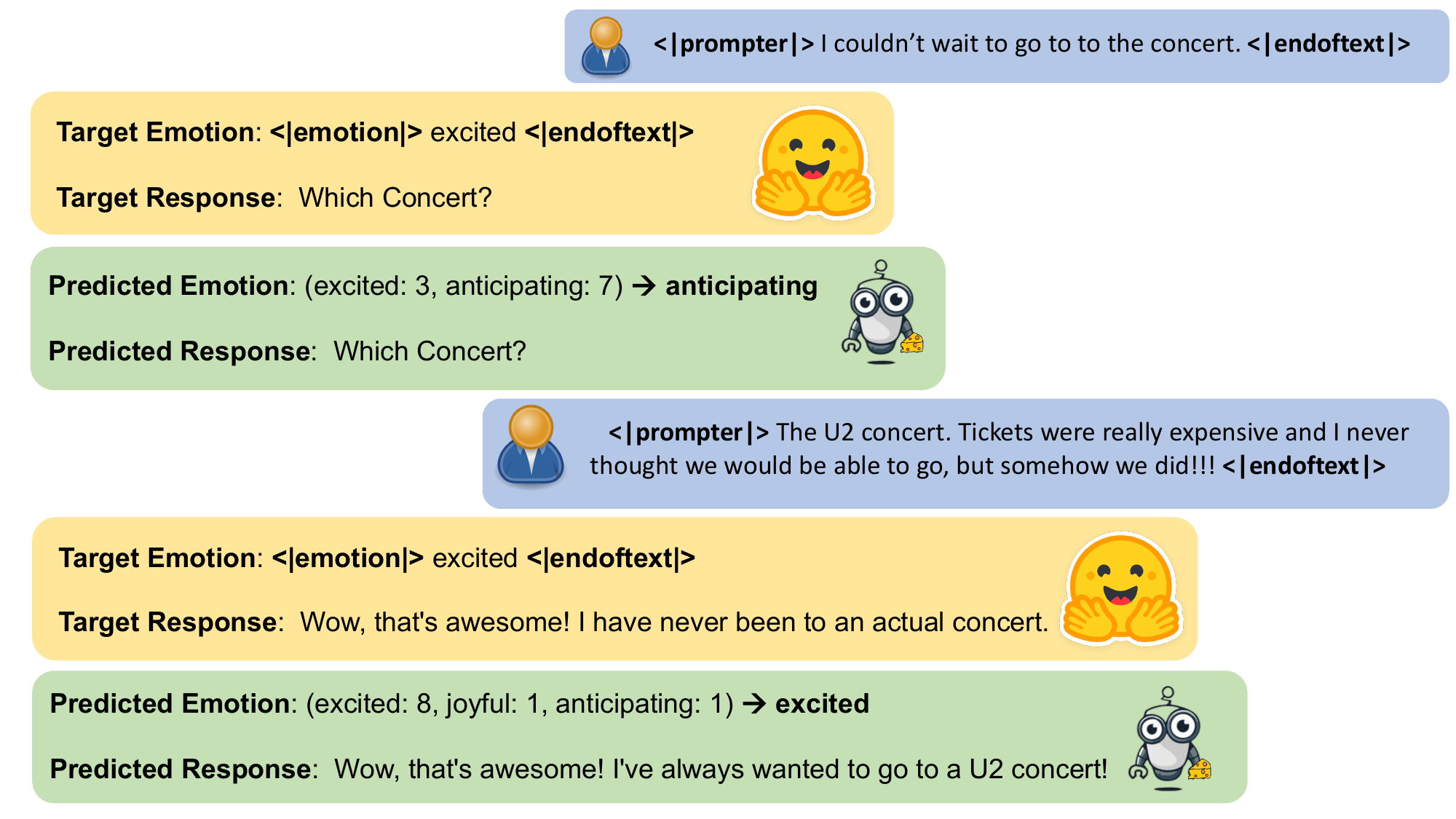}
    \caption{An example of how our 3-turns coversation structure has been implemented.}
    \label{convs}
\end{figure}
By recognizing and responding to a wide range of emotions, empathetic chatbots can tailor their responses to provide comfort, advice, or support, thereby enhancing the user experience. This capability is particularly valuable in applications ranging from customer service and mental health support to personal assistants and social companions, where understanding and addressing the emotional needs of users can significantly impact satisfaction and outcomes~\cite{crestani2022early,cena2023combining}.

With the advent of advanced large language models (LLMs), the interaction experience with conversational agents has seen remarkable improvements. 
These new models exhibit enhanced understanding of natural language, greater contextual awareness, and the ability to generate more coherent and contextually appropriate responses. 
This technological leap has not only transformed how conversational agents interact with users but has also opened new avenues for analyzing and understanding human emotional expressions~\cite{sekulic2021user}.

The motivation behind this research is rooted in the understanding that mental health support can be augmented through the use of empathetic conversational AI. For this purpose, we developed a conversational system called RACLETTE (\textit{\textbf{R}esponsive \textbf{A}nalysis with \textbf{C}hatbot \textbf{L}LMs for \textbf{E}motional and \textbf{T}herapeutical \textbf{T}racking and \textbf{E}valuation}). 

The key highlights of this paper include the development of a conversational model capable of detecting, understanding, and responding to emotional cues similar to human empathy (see Figure \ref{convs} for a visual example). This model is based on a novel approach to create emotion embeddings, which allows for the gradual construction of a user's emotional profile through interaction with the empathetic conversational model. We show how the user's emotional profile can be compared with known, pre-computed emotional profiles extracted from specialized datasets where individuals discuss their own experiences on specific mental health issues, with the rationale of potentially obtaining an explainable assessment of the mental state of the user engaging with the system.

The contributions and findings of this work are twofold: 
\textbf{(1.)} We define a method to tailor a chatbot, RACLETTE, for reacting empathetically to a specific user. RACLETTE uses an unconventional 3-turn structure where the model is trained to predict the user's emotion as a next-token prediction, leveraging the generative capabilities of the underlying Mistral 7B model, and responds empathetically based on the predicted emotion. Additionally, during the conversation, the user's emotional profile is updated, making the chatbot aware of the user's emotional condition in real-time. This updating allows the system to refine its understanding of the user's emotional state, enabling more precise and empathetic responses. 
\textbf{(2.)} We demonstrate how different mental disorders can be viewed as mixtures of specific emotions, guided by psychological theory. This framework suggests that emotional states are interconnected components forming distinct patterns linked to various mental health conditions. Emotional profiles for specific disorders can be pre-calculated and compared with users' emotional profiles to differentiate between conditions, potentially aiding in early detection and diagnosis. These emotional profiles can be viewed as markers for identifying different groups of mental disorders.

The paper is structured as follows. Section \ref{rel} discusses the Related Work. Section \ref{met} presents the methodology, data, and provides a psychological rationale supporting our approach. Section \ref{emo} presents the results of our model on the task of correct emotion classification and the quality of empathetic response generation. Section \ref{sec:reddit-embeddings} discusses the explainable method for generating embeddings associated with various mental disorders and shows qualitative results. In Section \ref{sec:reddit-suicide} we present the results of an experiment using emotional profiles to discriminate users belonging to different subreddit communities. Finally, Section \ref{conc} presents the conclusions.
 
\section{Related Works}\label{rel}
Significant research efforts have been devoted to developing sophisticated conversational models capable of understanding human emotions and generating empathic responses. The detection of sentiment and emotions has been recognized as crucial for the development of empathetic chatbots, as highlighted in~\cite{felbo2017,xu2018,shin2019,zhou2020}. These works underline the importance of integrating emotional understanding capabilities into automatic dialogue systems to enhance human-computer interaction. In~\cite{morris2018towards}, authors demonstrated the feasibility of using corpus-based approaches to enable conversational agents to simulate subtle empathy. 
Recent research has focused on developing personalized conversational systems that can maintain coherence and user engagement throughout interactions \cite{madotto2019personalizing,cho2022personalized}. These systems aim to create more natural and personalized dialogue experiences by adapting their responses to specific user characteristics and conversation contexts.

Furthermore, the comprehensive scoping review by~\cite{abd2021perceptions} sheds light on patient perceptions of mental health chatbots, revealing a positive outlook but emphasizing the need for enhanced linguistic capabilities and personalized interactions.

Recently, there has been a significant rise in the application of NLP techniques within the field of psychology~\cite{le2021machine}. This growing interest stems from the ability of NLP to extract valuable linguistic markers from both spoken and written communication, offering crucial insights into various mental health disorders~\cite{agurto2023speak,corcoran2020language,corona2023natural,HE2024115752}. 

Research has shown, for example, that measures of language coherence can serve as strong predictors of psychotic symptoms in individuals at high clinical risk~\cite{just2020modeling}. Clearer language production deficits are typically observed during the first episode of psychosis~\cite{gargano2022language}. One of the core symptoms, language disorganization, can be evaluated by analyzing the coherence and logical consistency of speech. For example, topic models~\cite{blei} have been used to assess psychotic symptoms during patient interviews. 
In this context, the use of markers proves valuable for identifying differences within patient groups in an interpretable way. Our method aligns with the trend of leveraging the representational power of large language models to create useful markers for identifying trends within populations. This approach extends the current research in NLP applications for mental health, where language patterns serve as indicators of psychological states. By using emotion embeddings as markers, our method offers a novel way to quantify and analyze the emotional content of language, offering a computational framework for understanding mental health through affective patterns. Similar to how language coherence and organization have been used to predict psychotic symptoms, these emotional markers could potentially serve as early indicators or diagnostic aids for a range of mental disorders. 

Building upon these foundational works, this study draws inspiration from CAiRE’s empathetic neural chatbot model by~\cite{lin2020caire}, and the innovative approach of using grayscale labels for emotion recognition as suggested by ``\textit{The Emotion is Not One-hot Encoding}'' by~\cite{lee2022emotion}.

\section{Methodology}\label{met}

This work proposes a novel methodology, guided by the intuition that one of the fundamental qualities of a therapist is \textit{empathy}. This direction aims to synthesize empathetic responses based on a broader understanding of affective language, circumventing the need for sensitive, real-world conversational data, enabling the model to detect emotions, and create explainable emotional profiles that can be useful for mental health assessment.

\subsection{A Psychological Rationale for our Approach}

Empathy has two main components:\\
    \textbf{(1.)} \textit{Cognitive Empathy}, the intellectual ability to understand another person's emotions, thoughts, and motives. It involves the ability to comprehend someone else's mental state and why they might be feeling a certain way, which is essential for effective communication and social interaction.\\
    \textbf{(2.)} \textit{Affective Empathy}, the ability to physically feel another person’s emotions, often leads to emotional responses such as compassion or concern. For a more detailed discussion, see \citep{decety2005perspective}.

\begin{figure}
    \centering
    \includegraphics[width=1.\linewidth]{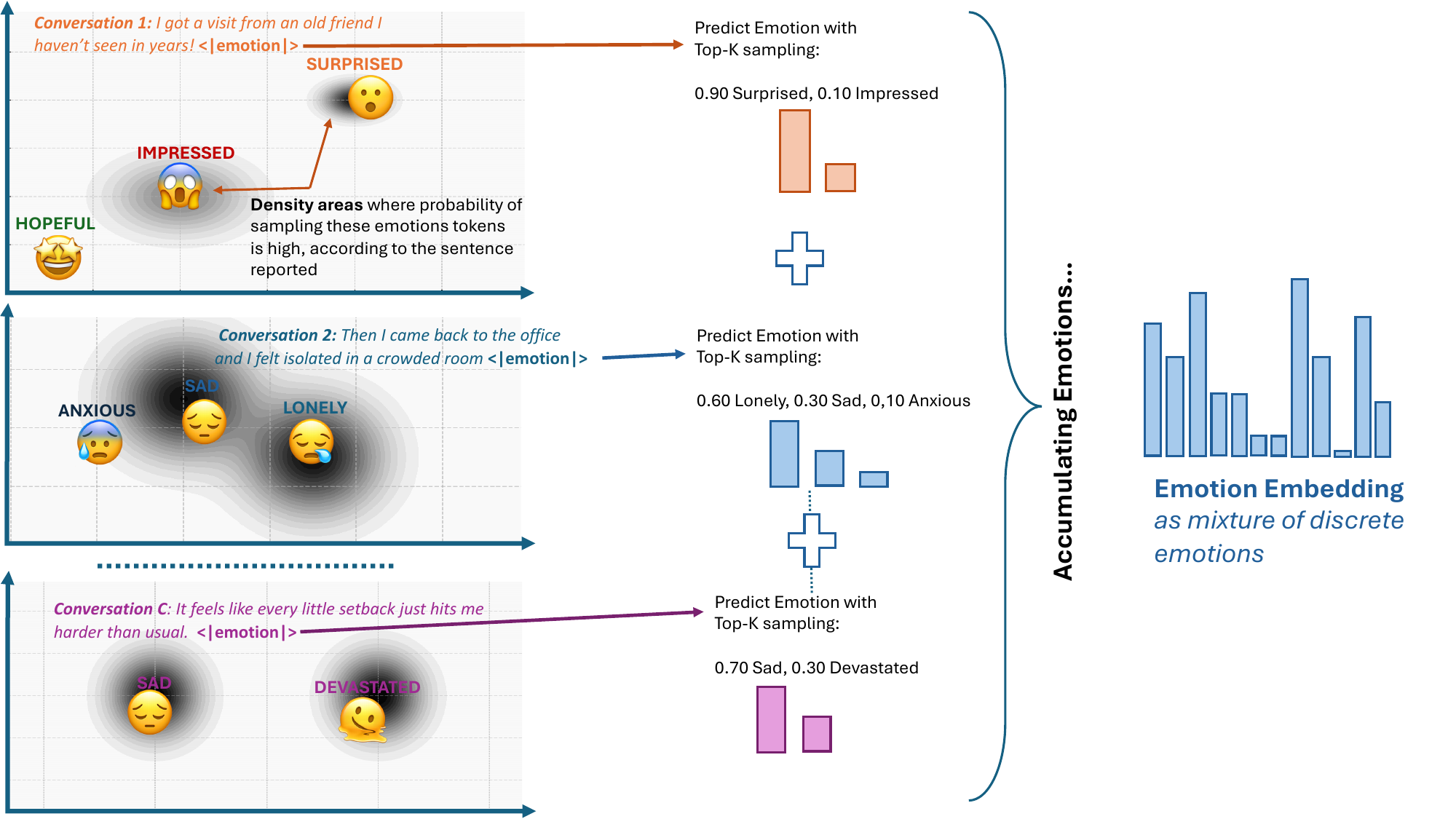}
    \caption{A visual explanation of how the emotional profile of a user is updated across a conversation and how to extract the final emotion embedding. }
    \label{convacc}
\end{figure}

This research focuses on Cognitive Empathy, aiming to classify the emotional state of a patient and enable the system to respond appropriately. Teaching machines to physically feel the emotions of others raises deep ethical and philosophical questions about the nature of consciousness and emotion in artificial systems, a topic that will likely remain at the forefront of futuristic research.

The approach aims to leverage and enhance the capabilities of empathetic LLMs by integrating emotion embeddings into their framework, guided by the intuition that the emotional spectrum is complex, and many emotions may coexist in a single sentence or piece of text.

We define an \textit{emotion embedding} as a high-dimensional vector representing an emotional state. Unlike word embeddings~\cite{allen2019analogies}, which capture semantic meaning, emotion embeddings synthesize an individual's emotional state within a conversation by encoding affective information. A distribution is generated by sampling and normalizing multiple predictions from a probabilistic classifier. These distributions can also be interpreted as embeddings, enabling meaningful algebraic operations. Complex emotions are encoded and represented by sequentially accumulating through the summation of many emotion embeddings, e.g., by accumulating over the many interactions that occur over an entire conversation (see Figure \ref{convacc} for a visual explanation).

An emotion embedding can be defined as: 

\[\text{Emotion Embedding} = \sum_{j}^{K} \alpha_j e_j \]

\noindent where \(\sum_{j}^{K} \alpha_j = 1\), \(K\) is the total number of all the different type of emotions considered and \(e_j\) is a specific emotion.

The use of emotional profiles to assess whether patients suffer from mental disorders is not entirely new in psychometrics. This approach aligns with established psychological assessment methods, such as the Beck Depression Inventory-II (BDI-II)~\cite{beck1996beck}, a widely used tool for measuring depression severity. The BDI-II includes items evaluating various emotional states and symptoms, like sadness, pessimism, guilt, agitation, irritability, and indecisiveness. Each item contributes to an overall score, aiding in the formulation of a final diagnosis.

For example, in the BDI-II, a patient might score high on sadness and pessimism, while moderate on guilty feelings and irritability.  
Similarly, our approach creates an emotional profile capturing the interplay of various emotions, offering a comprehensive view of an individual's mental state.

Thus, our work reveals a key insight: emotions act as indicators of deeper, complex mental states.

This study aims to demonstrate that mental states can be represented as collections of different emotions. Therefore, explainable emotion embeddings can be useful not only in identifying individuals in need of assistance but also as a potentially effective tool for automated diagnosis.

This multidimensional approach to emotional assessment acknowledges the complexity of human psychology, where a single emotional label cannot fully capture an individual's experience. By analyzing the emotional distribution, our system provides insights that align with the subtle understanding of mental states in clinical psychology, potentially enabling more accurate and personalized mental health support.

\subsection{Datasets}

In this work, three main sources of open-source data were used: \\
\noindent \textit{Empathetic Dialogues Dataset}~\cite{rashkin2019empathetic}: it has been used to train the RACLETTE model to identify emotions and respond empathically. This dataset is a large-scale multi-turn empathetic dialogue dataset collected on the Amazon Mechanical Turk, containing 24,850 one-to-one open-domain conversations. This dataset was selected for this task because, other than its high quality and appropriate size, it considers a much wider range of emotions compared to all other available datasets, which usually consider only a very limited subset (5-8) of fundamental emotions~\cite{zahiri2017emotion,li2017dailydialog}.\\
\noindent \textit{Reddit Mental Health Dataset}~\cite{low2020natural}: a collection of posts from specific Reddit forums (also called subreddits, Table \ref{combined} shows all the subreddits considered) have been used to construct the discrete distributions associated with each mental disorder to extract emotion embeddings.\\
\noindent\textit{DailyDialog Dataset}~\cite{li2017dailydialog}: a collection of posts used to establish a control group for the emotional profiles assessment. It contains 13,118 dialogues split into a training set with 11,118 dialogues and validation and test sets with 1,000 dialogues each.

\begin{figure}
    \centering
    \includegraphics[width=.8\linewidth]{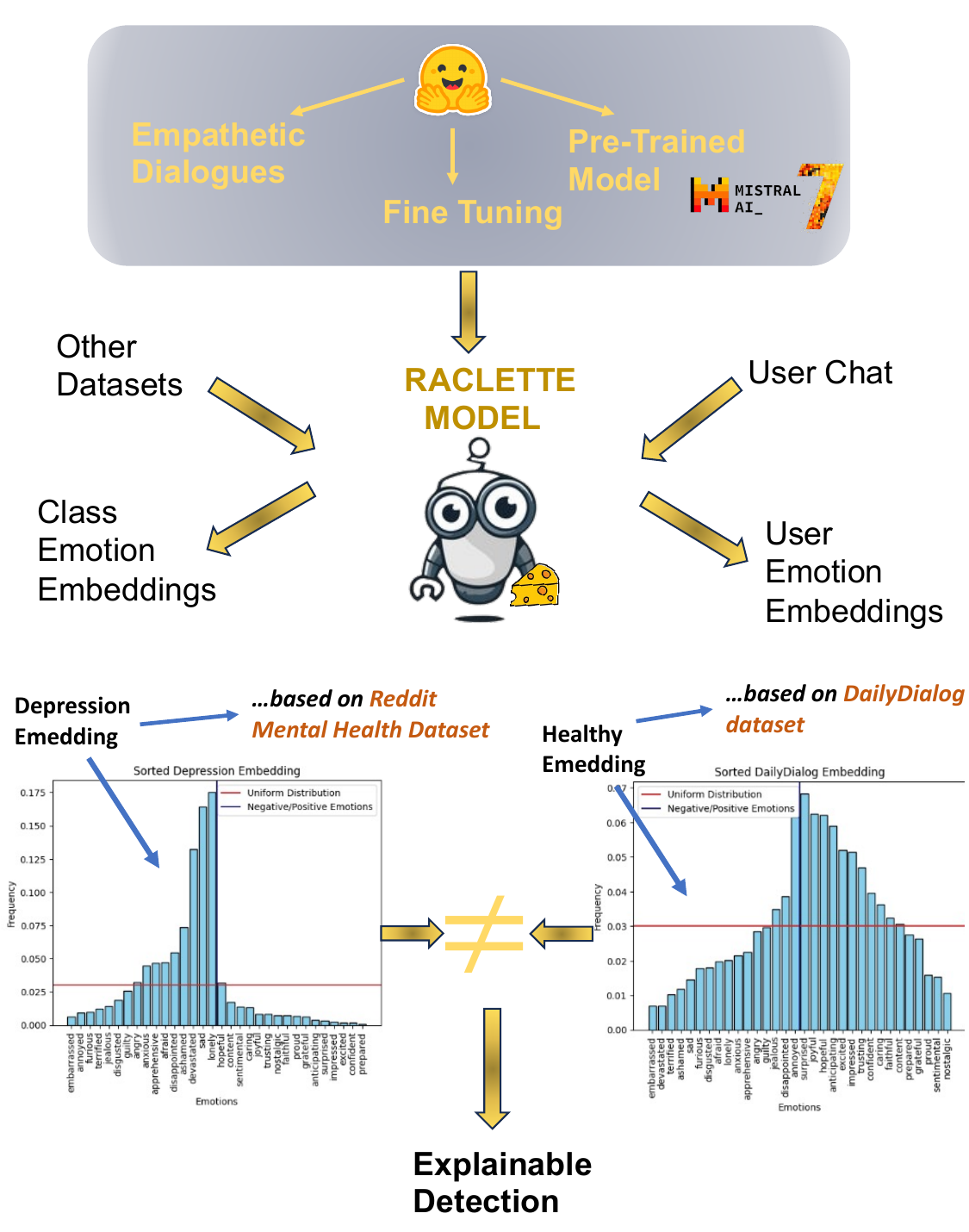}
    \caption{Overview of the main steps of RACLETTE pipeline.}
    \label{raclette_pipeline}
\end{figure}

Figure \ref{raclette_pipeline} illustrates the RACLETTE workflow pipeline, highlighting the specific use of each dataset at various stages of the process. This visual representation provides a clear overview of how the different datasets are integrated into the system's architecture, from training the empathetic model to extracting emotion embeddings and conducting comparative analyses.

\subsection{Tailoring an LLM to React Empathetically}

For this study, we chose to fine-tune the Mistral 7B model~\cite{jiang2023mistral}, a 7-billion-parameters state-of-the-art LLM, known for its great performance combined with both computational and memory efficiency. This approach aligns with recent findings from \cite{sekulic2024reliable}, which demonstrate that fine-tuning LLMs on task-oriented dialogue data can reduce hallucinations. 

The Empathetic Dialogues Dataset is purposefully formatted using a specific structure that allows to fully leverage the causal attention mask of the transformer decoder model to generate tokens for both the empathic response prediction and dialogue emotion detection tasks. Placing prompts before emotion labels enforces the autoregressive property of the model during training and inference~\cite{sun2023text}, allowing the generative model to be used both as a classifier and a conversational agent. 
The model learns to predict the next tokens by only attending to previous positions in the sequence in order to generate predictions sequentially.
Let \( P(y_{<emotion>} | y_1, y_2, \ldots, y_{N}) \) be the probability of the model predicting the emotion token \( y_{<emotion>} \) given the sequence of previous tokens \( y_1, y_2, \ldots, y_{N} \), then the model's objective can be defined as: $$\max \sum_{t=1}^{N} \log P(y_{<emotion>} | y_1, y_2, \ldots, y_{N})$$
where \( N \) is the length of the sequence.

When generating the prediction of an emotion, the model iteratively produces the tokens that are more likely, given the previous tokens (see Figure \ref{convacc}). Unlike deterministic methods, this process can be guided to generate a diverse set of emotions by iteratively sampling over the predicted probability distribution of all tokens in the vocabulary. In this implementation, Top-K Sampling is used~\cite{holtzman2019curious}, which limits the sampling pool to the top-K most probable tokens, in this case, \textit{top-10}, balancing diversity with relevance. Then to generate multiple emotions, this process is repeated \textit{10} times independently for each prompt. 
Let $V$ be the vocabulary and $K$ be the sampling parameter:
$$TopK(P(y_t | y_{<t}), K) = {y_i \in V : P(y_i | y_{<t })}$$ 
is among top-k probabilities. 

These empirical distributions are aggregated across the entire conversation to obtain the emotional profile of the speaker.
Let $C$ be the set of all prompts in a conversation and $e_{i,k}$ the sampled emotion ($K = 10$ samples in total) for prompt $i$ :
$$EmotionalProfile = \frac{1}{|C|} \sum_{i \in C} \frac{1}{K} \sum_{k=1}^K e_{i,k}$$

For this study, an unconventional 3-turns structure was implemented (see Figure \ref{convs} for an example). It can be summarized as Prompt, Emotion, and Response, separated by the special tokens: 
\(<|prompter|>,<|emotion|>,<|assistant|>\) and \(<endoftext>\).

When predicting empathetic responses, the model will attend to the previous tokens in its context, \textbf{(1)} the whole history of the conversation, \textbf{(2)} the current prompt followed by the emotion, and learn to generate the appropriate reply as seen in the training dataset.

\section{Emotion Recognition and Empathetic Response} \label{emo}

Table \ref{summ} shows the results of RACLETTE in detecting the correct emotion for each conversation both at prompt and conversation levels. 
The results related to individual prompts refer to the correct classification of emotions for single conversation turns. Regarding the conversation, this approach progressively concatenates each prompt, its predicted emotion, and subsequent response, thus enriching the model's contextual awareness with each conversational turn. The accumulated emotion distributions for each prompt contribute to a more precise classification, resulting in a 3\% increase in accuracy (from 56\% to 59\%). Notably, this methodology enhances accuracy, as the expanding conversational context provides more information for discerning the speaker's emotional profile.

Out of approximately 10.9k utterances present in the test set, the report categorizes 5,242, as the classification is made solely on the speaker's contributions.
In addition, to evaluate empathic replies to each of the speaker prompts, this table includes the BERTSCORE~\cite{zhang2019bertscore}, an automatic evaluation metric for text generation. Unlike traditional metrics that rely on exact word matches or n-grams, BERTSCORE evaluates the similarity between predicted and target replies by analyzing contextual embeddings of tokens obtained with the BERT model. This approach allows for a semantical understanding of the model's performance, capturing the comparison of empathetic responses beyond mere lexical matching. Notably, a BERTSCORE of 0.87
indicates high semantic similarity between the responses given by the model and the target replies contained in the test set that were given by the human listeners. 

Table \ref{tab:emotion_accuracy} compares the overall emotional accuracy of RACLETTE with the accuracy of CAiRE, as reported by~\cite{lin2020caire}. 
For completeness, in addition to CAiRE, we present other baselines from the literature that have used the same dataset to evaluate their performance.
As benchmarks for emotion classification accuracy, we consider the following approaches:
\textbf{(1.)} In \cite{chen2024cause}, authors propose a cause-aware empathetic generation method using Chain-of-Thought fine-tuning on Large Language Models. 
\textbf{(2.)} In \cite{li2022knowledge},  authors introduced a knowledge-enhanced empathetic dialogue generation method incorporating external knowledge and emotional signals. 
\textbf{(3.)} The approach from \cite{gao2021improving}, who proposed incorporating emotion cause recognition into empathetic response generation using an emotion reasoner and gated attention mechanism.

We report the accuracy value as presented in works that have used the same dataset, as shown in their respective manuscripts.

Table \ref{tab:emotion_accuracy} shows how RACLETTE  outperforms the benchmarks considered. 
Also in this case, it can be observed that the choice of fine-tuning a generative model, leveraging its autoregressive characteristics for classification, leads to the best results.

\begin{table}[ht]
\small
\centering
\resizebox{\columnwidth}{!}{%
\begin{tabular}{lcccc}
\hline
\textbf{Emotion} & \textbf{Precision} & \textbf{Recall} & \textbf{F1-Score} & \textbf{Support} \\ \hline
\multicolumn{5}{|c|}{\cellcolor{mygray}\textcolor{black}{Individual Prompts (5,242 Prompts)}} \\ \hline
Macro avg     & 0.56 & 0.56 & 0.55 & \\
Weighted avg  & 0.56 & 0.56 & 0.55 & \\ \hline
BERTSCORE   & 0.873      & 0.865   & 0.869     &  0.87        \\ \hline
Accuracy &  &      &      & 0.56 \\ \hline
\multicolumn{5}{|c|}{\cellcolor{mygray}\textcolor{black}{Conversations (2,472 Conversations)}} \\ \hline
Macro avg     & 0.59 & 0.59 & 0.58 & \\
Weighted avg  & 0.59 & 0.59 & 0.58 & \\ \hline
Accuracy &  &      &      & 0.59\\ \hline
\end{tabular}
}
\normalsize
\caption{Emotion Classification Report: Evaluated on \textit{Individual prompts} and \textit{Conversations} from the Empathetic Dialogues test set.}
\label{summ}
\end{table}

\begin{table}[ht]
\small
\centering
\begin{tabular}{|l|c|}
\hline
\textbf{Model} & \textbf{Emotional Accuracy} \\
\hline
\cite{gao2021improving}& 0.42 \\
\hline
\cite{li2022knowledge}& 0.46 \\
\hline
\cite{chen2024cause}& 0.53 \\
\hline
CAiRE & 0.51 \\
\hline
\textbf{RACLETTE } & \cellcolor{verdescuro}0.59 \\
\hline
\end{tabular}
\caption{Emotional accuracy comparison between RACLETTE and other benchmarks (best results highlighted in green). }
\label{tab:emotion_accuracy}
\normalsize
\end{table}

\section{Mental State as Mixtures of Emotions} 
\label{sec:reddit-embeddings}

This experiment shows a possible novel approach to create explainable mental state embeddings based on emotions, expanding the conversational model’s role from empathetic response generation to emotion analysis and diagnostic tool. The approach involves leveraging the fine-tuned model from the previous experiment, primarily as an emotion classifier. The idea is to extract emotional embeddings, used as \textit{markers} that are indicative of specific mental disorders from specialized corpora, in this case from social media interactions in mental health forums. The goal is to later compare the distinctive emotional profiles obtained in this experiment to the profiles obtained from users interacting with the model in a conversation.

For this experiment, 
considering the lack of professionally labeled data,  various datasets are gathered from Reddit, a social news website and forum where content is socially curated and promoted by site members through voting. It must be acknowledged that the data obtained from Reddit or other social media platforms may not accurately represent the broader population with mental illnesses, as it only captures those who choose to discuss their experiences online. 

Reddit is organized into 
forums known as ``subreddits''. Each subreddit focuses on a specific topic, interest, or theme, creating a unique community within the broader Reddit platform. In Table \ref{combined} all the considered subreddits obtained from~\cite{low2020natural} are reported (for a more in-depth discussion we refer to Section \ref{redd} in Appendix), together with a graphical visualization in Figures \ref{distributions} and \ref{distributions2} in Appendix of the relative mental state  embeddings based on emotions obtained by processing 1,000 posts from each subreddit. 

In the approach described in this section, embeddings for each mental disorders were generated by processing posts from the respective subreddits. The methodology involves the empathetic conversational model obtained in the previous experiment. However, rather than responding with both emotion and a reply, the posts are segmented into individual phrases. For each phrase, the model predicts a set of 10 emotions. These predicted emotions are then aggregated across all posts by summation and subsequently normalized. This process results in a characteristic emotional distribution profile for each mental disorder.

\begin{figure*}[ht]
    \centering
    \includegraphics[width=1.\linewidth]{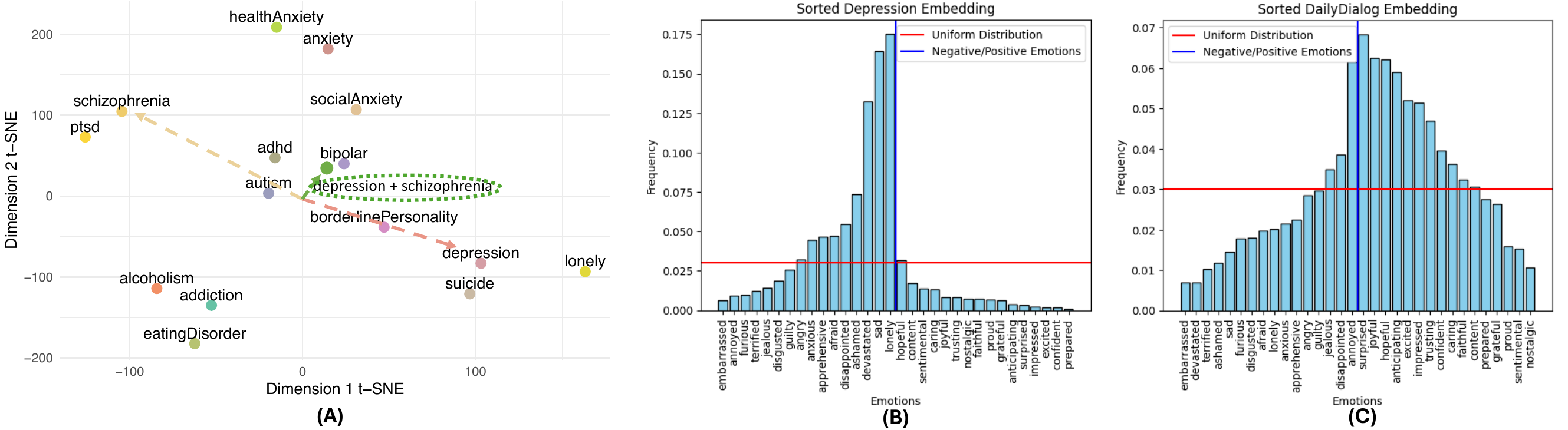}
    \caption{\textbf{(A)} 2-Dimensional representation of mental disorders distribution after applying t-SNE dimensionality reduction. \textbf{(B)} Sorted emotion embedding of depression. \textbf{(C.)} Sorted emotion embedding of DailyDialog.}
    \label{tsne}
\end{figure*}

This experiment yielded interesting results: the obtained emotion embeddings show significant differences across a spectrum of Reddit communities. Also, similarities across related disorders were to be expected. For instance, depression and suicide, or addiction and alcoholism,
show consistent similarities. Overall, these distributions can provide insights into how individuals discussing their experiences with similar conditions might perceive and express their feelings. In Figure \ref{tsne}\textbf{(A)}, we observe the mental disorder representations in a two-dimensional reduced space after applying t-SNE on top of the emotional embeddings. We can observe how mental disorders such as \textit{alcoholism}, \textit{addiction}, and \textit{eating disorder} are close in the embedding space, as are \textit{depression} and \textit{loneliness} or \textit{schizophrenia} and \textit{post-traumatic stress disorder (PTSD)}. We can also observe interesting properties of our representations, for example, by summing the depression and schizophrenia's embeddings, a new representation can be obtained, that is very close, in the embedding space, to bipolar. 

The assumption that mentally distressed individuals show identifiable skewed patterns of emotions must be addressed, by first establishing a normal distribution for comparison. For this, the Daily Dialogue Dataset~\cite{li2017dailydialog}, a high-quality multi-turn open-domain English dialogue dataset, was chosen as a control group. On average there are around 8 speakers turns per dialogue with around 15 tokens per turn where people discuss their daily lives, the whole training set was used to extract the embedding for this dataset.

The order of the emotional features in the embeddings can be arbitrary. But as an example, for clarity and visual comparison, the control group embeddings and the depression embeddings can be ordered according to what are most commonly considered positive and negative emotions: Figure \ref{tsne}\textbf{(B-C)} clearly shows the contrast in emotional profiles, emphasizing the marked disparities in how emotions are manifested and experienced by those within the reddit depression community, exhibiting an extremely skewed distribution towards negative emotions, compared to individuals engaging in daily dialogues.

\section{Reddit's Emotion embeddings applied to the Detection of Suicide Risk}
\label{sec:reddit-suicide}
This experiment aims to evaluate the use of emotion to create mental state embedding as a mechanism for diagnosing the potential risk of suicide. For lack of a professionally labeled dataset, this experiment, like the previous one, focuses on Reddit's users. 
To compare our approach with other related tasks and methods, we have built a dataset for binary classification of general conversation text versus suicidal text. We used the two subreddits CasualConversation and SuicideWatch, where CasualConversation is a subreddit for general conversation, and has generally been used by other methods as data for a clinically healthy class in other works~\cite{haque2021deep,shen2017detecting}. This dataset is part of a larger collection available on Kaggle\footnote{\href{https://www.kaggle.com/datasets/nikhileswarkomati/suicide-watch}{ https://www.kaggle.com/datasets/suicide-watch }}, which has been carefully cleaned to ensure the reliability of the data. We select 5\% of the initial samples at random as a test set  ($\approx$ 10,585 samples). After inspecting posts for anomalous length deviations from the average, those lacking informative content are removed.

To compute the sample embeddings, each post is divided into sentences, with 10 emotions predicted per sentence. The final embedding aggregates these emotions across all sentences in the post, following a similar approach as used in Section \ref{sec:reddit-embeddings} to extract reference embeddings. 

The methodology encompasses the use of three distinct metrics for comparing embeddings: Kullback–Leibler (KL) divergence, Jensen–Shannon (JS) divergence, and Cosine Similarity (CS).
Our approach is based on the use of emotional profiles embeddings that are most closely associated with an elevated risk of suicide and match them against the user's emotional profile. 
Table \ref{combined} compares various emotion embeddings, focusing on their differences from the suicide embedding, measured by KL, JS, and CS. The results show emotional proximity between Suicide and Depression, as well as Borderline Personality Disorder, followed by Bipolar Disorder, Addiction, PTSD, and Schizophrenia. This pattern aligns with psychological insights that these mental health conditions are often linked to an increased risk of suicidal tendencies \citep{Song2020ComparisonOS}.
\begin{table}[ht]
\small
\centering
\begin{tabular}{|l|c|c|c|}
\hline
\textbf{Emotion} & \textbf{KL} & \textbf{JS} & \textbf{CS} \\
\hline
suicide         & 0.000 & 0.000 & 1.000 \\
\hline
depression      & 0.062 & 0.124 & 0.969 \\
\hline
bpd             & 0.201 & 0.226 & 0.852 \\
\hline
bipolar         & 0.451 & 0.332 & 0.637 \\
\hline
addiction       & 0.519 & 0.349 & 0.592 \\
\hline
ptsd            & 0.545 & 0.355 & 0.594 \\
\hline
alcoholism      & 0.567 & 0.355 & 0.586 \\
\hline
schizophrenia   & 0.745 & 0.407 & 0.519 \\
\hline
eatingDisorder  & 0.822 & 0.409 & 0.496 \\
\hline
socialAnxiety   & 0.830 & 0.432 & 0.463 \\
\hline
uniform         & 0.850 & 0.449 & 0.540 \\
\hline
autism          & 0.855 & 0.435 & 0.467 \\
\hline
adhd            & 1.016 & 0.454 & 0.428 \\
\hline
anxiety         & inf   & 0.481 & 0.303 \\
\hline
lonely          & inf   & 0.429 & 0.579 \\
\hline
healthAnxiety   & inf   & 0.591 & 0.260 \\
\hline
\end{tabular}
\caption{ Comparison of Emotion Embeddings: KL Divergence (KL), JS Divergence (JS), and Cosine Similarity (CS) w.r.t. suicide embedding.}
\label{combined}
\normalsize
\end{table}

For these reasons, this experiment will focus on the use of these specific emotion embeddings, considering the embedding that is most similar to what is obtained by processing the sample post, and mapping it to the predicted label, in an unsupervised fashion, as follows:

\textbf{Positive labels:} suicide, depression, borderline personality disorder (BPD), bipolar disorder, PTSD, addiction, and schizophrenia. By combining these particular embeddings, the study aims to capture a spectrum of characteristic emotional patterns that are potentially indicative of an elevated risk of suicide.

\textbf{Negative labels:} normal and uniform distributions, where normal is obtained from the Daily Dialogue dataset, and embedding obtained from Casual Conversation's subreddit are also used as the control group.

\subsection{Results for Mental Health Classification} 

Table \ref{tab:results} shows the performance metrics: precision, recall, F1 score and accuracy. For each similarity metrics — Kullback–Leibler divergence, Jensen–Shannon divergence, and Cosine Similarity — we also introduce a combined method, where if any of these methods detect a risk of suicide, the label is assigned as positive. This experiment is designed to maximize recall, a critical metric in scenarios where missing a positive instance has severe consequences, thus reducing the emphasis on precision and false positives. As the quantitative results show, this approach achieves high recall at the cost of other metrics. 
\begin{table}[ht]
\small
\centering
\begin{tabular}{lrrrr}
\toprule
      Models &  Prec &  Rec &  F1 &  ACC \\ \hline
\multicolumn{5}{|c|}{\cellcolor{mygray}\textcolor{black}{RACLETTE}} \\ \hline
    KL Divergence &  \cellcolor{verdechiaro}  0.71 &  0.90 &  \cellcolor{verdescuro}   0.79 & \cellcolor{verdescuro}  0.77 \\ \hline
            JS Divergence  &     0.67 & \cellcolor{verdechiaro} 0.93 &  \cellcolor{verdechiaro}  0.78 &  \cellcolor{verdechiaro}  0.76 \\ \hline
             Cosine Similarity &     0.65 &\cellcolor{verdechiaro} 0.93 &    0.77 &    0.74 \\ \hline
            Combined &     0.63 & \cellcolor{verdescuro}  0.95 &    0.76 &    0.72 \\ \hline
\multicolumn{5}{|c|}{\cellcolor{mygray}\textcolor{black}{Benchmark Models}} \\ \hline
RoBERTa + KM & \cellcolor{verdescuro}  0.72 & 0.84 & \cellcolor{verdechiaro} 0.78 & \cellcolor{verdescuro} 0.77\\ \hline
BERT + KM &  0.65 &   0.80   &   0.71 &  0.69\\ 
\hline
\end{tabular}
\normalsize
\caption{Classification results for different methods of comparing embeddings to detect risk of suicide. The higher the score (a.k.a. the greener), the better.}
\label{tab:results}
\end{table}
Additionally, these methods are compared with state-of-the-art unsupervised approaches based on RoBERTa~\cite{liu2019roberta} and BERT's~\cite{devlin2019bertpretrainingdeepbidirectional} embedding representations, before grouping them into two classes using a K-Means clustering approach, as done in~\cite{subakti2022performance}.

The results show RACLETTE's Combined method achieving the highest recall of 0.95, indicating superior ability to identify relevant cases, though this comes with a trade-off in precision at 0.63. Conversely, RoBERTa leads in precision at 0.72, but with lower recall at 0.84. The KL Divergence variant of RACLETTE stands out for its balanced performance, maintaining strong scores across all metrics (precision: 0.71, recall: 0.90, F1: 0.79, accuracy: 0.77). Both JS Divergence and Cosine Similarity methods show similar patterns, with high recall (0.93) but lower precision.  The color intensity (green shading) in Table \ref{tab:results} indicates better performance, visually highlighting that RACLETTE's approaches generally outperform the benchmark models.

A key advantage of this method is that it generates explainable representations and emotion embeddings, which can be visually inspected, providing valuable insights into an individual's emotional profile.

\section{Conclusions} \label{conc}

This paper introduces the RACLETTE system, which addresses two critical challenges in mental health support: the need for empathetic AI-driven conversation and reliable assessment tools. By fine-tuning a LLM, we demonstrate that it's possible to create effective conversational agents that can accurately recognize users' emotional states while generating high-quality empathetic responses, all while avoiding the use of sensitive clinical data. The system not only achieves state-of-the-art performance in emotion recognition, but also introduces a novel methodology for creating emotional profiles. These profiles, generated by aggregating emotion distributions from user interactions, serve as interpretable markers that can be compared with characteristic patterns associated with various mental health conditions. Our experimental results demonstrate both the system's effectiveness in maintaining empathetic conversations and its potential as a preliminary screening tool through the analysis of emotional embeddings.

\section{Limitations}

One of the main critical point of this work is represented by the quality and reliability of the emotional data used for training. Emotional data must be diverse and accurately labeled to ensure the model can understand and respond to a wide range of emotional expressions. This data collection process is complex and time-consuming, often requiring manual annotation by experts to maintain high standards.

Furthermore, findings have revealed that individuals affected by mental disorders commonly turn to social media to share their personal experiences, seek out information about mental health and treatment options, and either offer or gain support from others who are dealing with similar challenges~\cite{naslund2020social,dodemaide2022social}. However, noise in the data is another significant limitation. For example, individuals may seek advice on behalf of others, such as family members, which can introduce inaccuracies. Self-reported information, while valuable, may not always be as reliable or accurate as clinically diagnosed conditions due to personal biases, misunderstandings, or intentional misreporting. Additionally, online self-expression can vary greatly between individuals, influenced by factors such as cultural differences, personal communication styles, and the specific context of the interaction.

Confounding factors, such as comorbidities, must also be taken into account. Individuals with multiple overlapping conditions may exhibit complex emotional and psychological profiles that are difficult for the model to parse accurately. Moreover, the way individuals express themselves online can differ significantly from in-person interactions, adding another layer of complexity to the model’s ability to interpret and respond appropriately.

Despite these challenges, the methodology addresses crucial privacy and confidentiality issues that are particularly important in the mental health domain. However, it does not fully address the ethical implications of using AI as a clinical tool, including the potential for misuse and the need for safeguards against harmful or biased behaviors in the conversational model. Continuous improvements and validation against clinical standards are essential to ensure that these tools effectively integrate into traditional care pathways, enhancing rather than disrupting the therapeutic process.

\section{Ethical Considerations}

The proposed methodology for mental health support and assessment, while innovative, brings several ethical considerations to the forefront that must be addressed to ensure responsible deployment.

There is a potential for AI to be misused as a clinical tool. Without proper safeguards, these models could exhibit harmful or biased behaviors, leading to adverse outcomes for users. 

Implementing ethical safeguards is crucial to mitigate the risks associated with AI in mental health. Developing clear guidelines on the appropriate use of AI, managing sensitive data protocols, and ensuring transparency in operations are essential steps. Involving ethicists, and clinicians in the development process will help create a balanced and ethical approach.

It is also crucial to clearly communicate the supplementary nature of these tools and the necessity of professional evaluation and treatment. There is a risk that users may become overly reliant on automated mental health support, potentially neglecting the importance of seeking help from qualified professionals. By ensuring that these tools are integrated into traditional care pathways, they can enhance the therapeutic process, providing additional support while maintaining the central role of professional mental health care providers.

\bibliography{custom}

\appendix
\section{Finetuning Details}
\label{finetuning_det}
For finetuning, this study employed SFTTrainer and QLoRa, implemented in the respective HuggingFace libraries~\cite{wolf2019huggingface,dettmers2023qlora}. The model parameters are quantized to the 4-bit NormalFloat(nf-4) datatype and the computations are performed in 16-bit BrainFloat (bFloat16). For reproducibility purposes, the following LoRa hyperparameters were used: scaling factor $lora\_alpha=16$, dropout probability $lora\_dropout=0.1$ and the rank of the update matrices $lora\_r=64$. The training hyperparameters: $batch\_size=1$, gradient accumulation $steps=16$, $warmup\_ratio=0.3$, cosine learning rate scheduler with an initial $l\_r=2e-5$, the model was trained for 3 epochs, using AdamW optimizer.

\section{Error Analysis}

This section presents the model's performance in predicting correct emotions from the Empathetic Dialogues dataset, analyzing results at both individual prompt level (Table \ref{tab:classification_report_sample}) and conversation level (Table \ref{tab:classification_report_conv}). While our model demonstrates overall good emotional accuracy, certain metrics for specific emotions exhibit suboptimal performance. This is particularly evident with emotions that are closely related but vary in intensity, such as `angry' and `furious'. These emotions, while technically distinct, can be challenging to differentiate even in human evaluation, as they often share similar underlying sentiments and can sometimes be considered interchangeable.
\begin{table}
\small
\centering
\begin{tabular}{lcccc}
\hline
\textbf{Emotion} & \textbf{Precision} & \textbf{Recall} & \textbf{F1-Score} & \textbf{Support} \\ \hline
afraid        & 0.45 & 0.23 & 0.30 & 152 \\ 
angry         & 0.29 & 0.25 & 0.27 & 170 \\
annoyed       & 0.61 & 0.63 & 0.62 & 186 \\ 
anticipating  & 0.43 & 0.32 & 0.37 & 152 \\ 
anxious       & 0.50 & 0.48 & 0.49 & 159 \\ 
apprehensive  & 0.42 & 0.40 & 0.41 & 146 \\ 
ashamed       & 0.47 & 0.33 & 0.39 & 135 \\ 
caring        & 0.62 & 0.71 & 0.66 & 164 \\ 
confident     & 0.53 & 0.52 & 0.53 & 156 \\ 
content       & 0.61 & 0.65 & 0.63 & 162 \\ 
devastated    & 0.48 & 0.57 & 0.52 & 139 \\ 
disappointed  & 0.56 & 0.53 & 0.54 & 165 \\ 
disgusted     & 0.72 & 0.75 & 0.73 & 175 \\ 
embarrassed   & 0.82 & 0.76 & 0.78 & 164 \\ 
excited       & 0.43 & 0.37 & 0.39 & 186 \\ 
faithful      & 0.71 & 0.72 & 0.71 & 103 \\ 
furious       & 0.42 & 0.33 & 0.37 & 141 \\ 
grateful      & 0.64 & 0.67 & 0.65 & 203 \\ 
guilty        & 0.61 & 0.72 & 0.66 & 135 \\ 
hopeful       & 0.50 & 0.52 & 0.51 & 163 \\ 
impressed     & 0.57 & 0.70 & 0.63 & 165 \\ 
jealous       & 0.88 & 0.75 & 0.81 & 167 \\ 
joyful        & 0.24 & 0.31 & 0.27 & 168 \\ 
lonely        & 0.73 & 0.86 & 0.79 & 159 \\ 
nostalgic     & 0.61 & 0.77 & 0.68 & 159 \\ 
prepared      & 0.64 & 0.71 & 0.67 & 157 \\
proud         & 0.57 & 0.60 & 0.59 & 200 \\ 
sad           & 0.46 & 0.40 & 0.43 & 179 \\ 
sentimental   & 0.68 & 0.36 & 0.47 & 189 \\
surprised     & 0.64 & 0.64 & 0.64 & 266 \\ 
terrified     & 0.42 & 0.72 & 0.53 & 143 \\ 
trusting      & 0.60 & 0.63 & 0.62 & 134 \\ 
\hline
TOTAL         &      &      &      & 5242 \\ \hline
\end{tabular}
\caption{Emotion Classification Report: Evaluated on \textbf{individual prompts} from the Empathetic Dialogues test set.}
\label{tab:classification_report_sample}
\end{table}

\begin{table}[ht]
\small
\centering
\begin{tabular}{lcccc}
\hline
\textbf{Emotion} & \textbf{Precision} & \textbf{Recall} & \textbf{F1-Score} & \textbf{Support} \\ \hline
afraid        & 0.49 & 0.24 & 0.32 & 70  \\ 
angry         & 0.33 & 0.27 & 0.30 & 82  \\
annoyed       & 0.64 & 0.68 & 0.66 & 88  \\ 
anticipating  & 0.47 & 0.32 & 0.38 & 69  \\ 
anxious       & 0.51 & 0.49 & 0.50 & 76  \\ 
apprehensive  & 0.46 & 0.45 & 0.45 & 67  \\ 
ashamed       & 0.52 & 0.35 & 0.42 & 63  \\ 
caring        & 0.64 & 0.70 & 0.67 & 77  \\ 
confident     & 0.59 & 0.57 & 0.58 & 70  \\ 
content       & 0.62 & 0.68 & 0.65 & 74  \\ 
devastated    & 0.51 & 0.56 & 0.54 & 66  \\ 
disappointed  & 0.66 & 0.59 & 0.62 & 81  \\ 
disgusted     & 0.72 & 0.81 & 0.76 & 84  \\ 
embarrassed   & 0.82 & 0.81 & 0.82 & 80  \\ 
excited       & 0.45 & 0.42 & 0.43 & 89  \\ 
faithful      & 0.78 & 0.72 & 0.75 & 50  \\ 
furious       & 0.44 & 0.33 & 0.38 & 67  \\ 
grateful      & 0.64 & 0.69 & 0.66 & 91  \\ 
guilty        & 0.61 & 0.75 & 0.67 & 63  \\ 
hopeful       & 0.59 & 0.58 & 0.58 & 78  \\ 
impressed     & 0.60 & 0.69 & 0.64 & 81  \\ 
jealous       & 0.89 & 0.82 & 0.85 & 78  \\ 
joyful        & 0.25 & 0.27 & 0.26 & 81  \\ 
lonely        & 0.76 & 0.88 & 0.81 & 75  \\ 
nostalgic     & 0.62 & 0.81 & 0.71 & 74  \\ 
prepared      & 0.68 & 0.77 & 0.72 & 75  \\
proud         & 0.60 & 0.67 & 0.63 & 95  \\ 
sad           & 0.51 & 0.48 & 0.49 & 86  \\ 
sentimental   & 0.72 & 0.36 & 0.48 & 87  \\
surprised     & 0.66 & 0.68 & 0.67 & 124 \\ 
terrified     & 0.43 & 0.73 & 0.54 & 71  \\ 
trusting      & 0.58 & 0.63 & 0.61 & 60  \\ \hline
TOTAL         &      &      &      & 2472\\ \hline
\end{tabular}

\caption{Emotion Classification Report: Evaluated on \textbf{conversations} from the Empathetic Dialogues test set.}
\label{tab:classification_report_conv}
\end{table}

\section{Beyond the 32 Emotion Classes}

Even though having 32 emotion classes may seem difficult enough for the classification task, ideally an empathic conversational agent should be able to understand and recognize the broadest possible range of emotions. 
Table \ref{tabne} shows that RACLETTE also correctly predicts emotions that are not part of the dataset used for fine-tuning, especially when prompts contain explicit references to these new emotions, showing a great understanding of the task. This phenomenon is indicative of our model's ability to generalize beyond its explicitly taught categories, showing that the model has effectively generalized the concept of emotion beyond its training examples, which is particularly fascinating in the context of emotion recognition. 
The base model already had some semantic understanding of the words associated with the concept of emotion, which are likely used in similar contexts and are similar to each other in the input embedding space. The fine-tuning process further enforced the similarity in the learned representations, while the generative method used for the classification task allows for more flexibility compared to conventional classification approaches.

In summary, this feature is a consequence of an unconventional use of a generative pre-trained transformer decoder model as a classifier. It allows the fine-tuned model to sometimes ``think outside the box'' of the constrained range of emotions typical of conventional classification approaches (Table \ref{tabne} shows examples of ``\textit{out-of-training-labels}'' predictions). For consistency in the experiments, these new emotions are discarded as outliers but serve as an interesting example to show the potential of generative models used for classification tasks.

\begin{table*}[ht]
\small
\centering
\begin{tabular}{>{\raggedright\arraybackslash}p{4cm} >{\raggedright\arraybackslash}p{4cm} >{\raggedright\arraybackslash}p{4cm}}
\toprule
\cellcolor{mygray}\textbf{Dialogue 1} & \cellcolor{mygray}\textbf{Dialogue 2} & \cellcolor{mygray}\textbf{Dialogue 3} \\
\midrule
\textbf{Prompt:}  I have four hours of boredom I'm currently bored out of mind waiting for ferry and then 2 hours on the ferry. Can anyone recommend mobile or two long films? 
& \textbf{Prompt:} I can’t feel anything, I feel detached to everyone and everything (even my mom and my sister). & \textbf{Prompt:} I’m so confused and barely even know what’s real anymore. \\
\textbf{New emotion:} bored & \textbf{New emotion:} disconnected & \textbf{New emotion:} confused \\
\midrule
\cellcolor{mygray}\textbf{Dialogue 4} & \cellcolor{mygray}\textbf{Dialogue 5} & \cellcolor{mygray}\textbf{Dialogue 6} \\
\midrule
\textbf{Prompt:} wondering why frisbees get bigger as they get closer. & \textbf{Prompt:} Then I get trapped into multiple situations, that I see no escape. & \textbf{Prompt:} If you don’t have an age flair You are hella sus. \\
\textbf{New emotion:} confused & \textbf{New emotions:} trapped & \textbf{New emotions:} suspicious \\
\bottomrule
\end{tabular}
\caption{Examples of dialogues in which new emotions, not present in the training dataset, are predicted by our empathetic model. }
\label{tabne}
\normalsize
\end{table*}

\section{Qualitative results and analysis}
\label{exampless}
The following examples show the inputs, targets, and prediction outputs, qualitatively comparing the predictions with the targets used to evaluate the model on the Empathetic Dialogues dataset. These example conversations are taken from the test set, during the quantitative evaluation of the fine-tuned model.
\\

\subsection{Dealing with coexisting emotions}
This example aims to show how the model deals with coexisting emotions, adjusting its prediction as the conversation goes along:
\begin{dialogue}
    \speak{Input Prompt} <|prompter|>I couldn't wait to go to the concert.<|endoftext|><|emotion|>
    \speak{Target Emotion}  excited
    \speak{Target Response}  Which Concert?
    \speak{Predicted Emotion} (excited: 3, anticipating: 7)$\rightarrow$ anticipating
    \speak{Predicted Response}  What concert was it?
\end{dialogue}

For the first prompt in this example conversation, the target emotion is `excited', and it will be the same for the whole conversation, as the dataset is structured this way. The 10 emotions predicted are 3 ``excited'' and 7 ``anticipating''. Given the prompt: ``I couldn't wait to go to the concert.'', both emotions classified by the model seem to be correct. Even though the target is ``excited'' the model decided that ``anticipating'' is more likely, which actually seem appropriate, if not more accurate than the target emotion.

Regarding the response, the model predicts a very similar response as the target, semantically the same.  The conversation goes on, appending the predicted emotion to the conversation history, during evaluation the target response is appended instead of the predicted response, otherwise, the conversations might not make sense. Because, obviously, the following prompt in the dataset is responding to the target reply of the previous one. So the conversation continues:
\\

\begin{dialogue}
    \speak{Input Prompt}\textbf{<|prompter|>}I couldn't wait to go to the concert.\textbf{<|endoftext|>}\\\textbf{<|emotion|>}anticipating\textbf{<|endoftext|>}\\ \textbf{<|assistant|>}What concert was it?\textbf{<|endoftext|>}\\\textbf{<|prompter|>}The U2 concert. Tickets were really expensive and I never thought we would be able to go, but somehow we did!!!\textbf{<|endoftext|>}
    \speak{Target Emotion}  excited
    \speak{Target Response}  Wow, that's awesome! I have never been to an actual concert.
    \speak{Predicted Emotion} (excited: 8, joyful: 1, anticipating: 1)$\rightarrow$ excited 
    \speak{Predicted Response}   Wow, that's awesome! I've always wanted to go to a U2 concert!
\end{dialogue}

 The final prediction is computed by aggregating all the emotions across the entire conversation, establishing `exited' as the prominent emotion, in this example:
\begin{dialogue}
    \speak{CONVERSATION TARGET EMOTION} excited
    \speak{CONVERSATION PREDICTED EMOTIONS} (`excited': 11, `anticipating': 8, `joyful': 1)$\rightarrow$ excited
\end{dialogue}

This example also provides an interesting insight on empathy, occasionaly the model seems to mimic the emotions of the prompter, in this conversation, the speaker expresses excitement and the model reacts by also expressing excitement. This behavior has been inferred from the data by imitating humans. Sometimes when emotions resonate to others, they become contagious, and the involountary empathic response is to feel and express this shared emotion, it is a natural way of showing understanding of how the other is feeling and helps people relate to each other. It is notably interesting to observe the model pick up on this subtle but fundamental aspect of empathy. \\

\subsection{Dealing with different intensities of the same emotion}
The following is another simple example from the evaluation on the test set, in this case, the response from the model is qualitatively more empathic than the target response. The model correctly identifies that the prompter is afraid, and more likely terrified, a correct prediction. It also responds accordingly, by acknowledging the implied feeling and its intensity:
\\

\begin{dialogue}
    \speak{Input Prompt}\textbf{<|prompter|>}I'm so scared of bugs! I found one in my hair yesterday and almost died.\textbf{<|endoftext|><|emotion|>}
    \speak{Target Emotion} terrified
    \speak{Target Response} ok
    \speak{Predicted Emotion} (afraid: 3, terrified: 7)$\rightarrow$ terrified
    \speak{Predicted Response} Oh no! That's so scary! What kind of bug was it?
\end{dialogue}

\subsection{Dealing with uncertainty}
The following example shows how one prompt is usually not enough to assess the emotional state of the speaker, it might be too short or ambiguous. The way the 10 predicted emotions are distributed across the 32 possibilities can provide insight into the uncertainty of the model: 

(The structure of the prompts and the conversation history are omitted for clarity)
\begin{dialogue}
    \speak{Input Prompt}The other day I had a deep craving for Chik-Fil-A.
    \speak{Target Emotion}  disappointed
    \speak{Target Response}  I understand that one,they are my favorite place to eat
    \speak{Predicted Emotion} ('disappointed': 2, 'content': 2, 'anticipating': 2, 'jealous': 1, 'disgusted': 2, 'hopeful':1)$\rightarrow$ disappointed
    \speak{Predicted Response} I love Chik-Fil-A! What did you get?
\end{dialogue}

Without more context, it is impossible to identify if the prompter was content, anticipating, disappointed, or maybe disgusted about having an intense craving for fast food. The model predictions reflect this uncertainty. Many different emotions are predicted, with a low count, so the overall prediction was a lucky one, as in the case of a tie, the first
emotion inserted in the list is selected. The conversation continues:

\begin{dialogue}
    \speak{Input Prompt}Mine too. So I drive there and pull in the parking lot and it's completely empty. I then realized it was Sunday.
    \speak{Target Emotion}  disappointed
    \speak{Target Response}  That has happened to me too many times to count
    \speak{Predicted Emotion} ('disappointed': 9, 'sad': 1)$\rightarrow$ disappointed
    \speak{Predicted Response} Oh no! That's a bummer. What did you end up getting instead?
\end{dialogue}

Given more context, it becomes clear that the speaker is disappointed about not getting the food he wanted. And the model prediction adapts to that:
\begin{dialogue}
    \speak{CONVERSATION TARGET EMOTION} disappointed
    \speak{CONVERSATION PREDICTED EMOTIONS} ('disappointed': 11, 'content': 2, 'anticipating': 2, 'jealous': 1, 'disgusted': 2, 'hopeful':1, 'sad': 1)$\rightarrow$ disappointed
\end{dialogue}

Even though uncertainty introduces noise, the correct emotion has the highest weight by far, thus it is correctly selected as the final prediction for this conversation.
\\

\section{Reddit's Subreddits}
\label{redd}

Following we discuss the reddit communities whose issues are related to the risk factors for suicide:\\

\begin{figure*}
    \centering
    \begin{subfigure}[c]{0.49\textwidth}
        \centering
        \includegraphics[width=\textwidth]{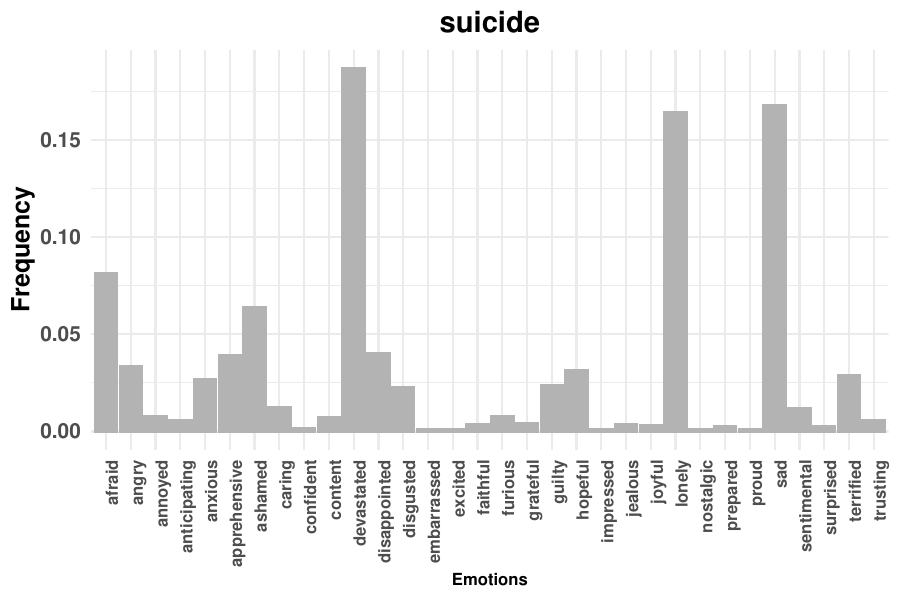}
        \caption{r/suicidewatch (1000 samples)}
        \label{fig:suicide}
    \end{subfigure}
    \hfill
    \begin{subfigure}[c]{0.49\textwidth}
        \centering
        \includegraphics[width=\textwidth]{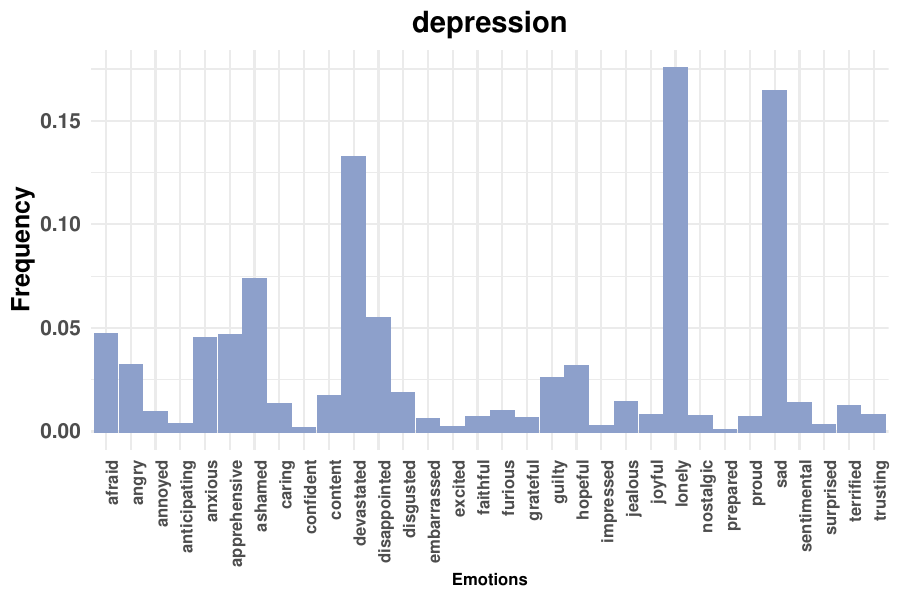}
        \caption{r/depression (1000 samples)}
        \label{fig:depression}
    \end{subfigure}
    \hfill
    \begin{subfigure}[c]{0.49\textwidth}
        \centering
        \includegraphics[width=\textwidth]{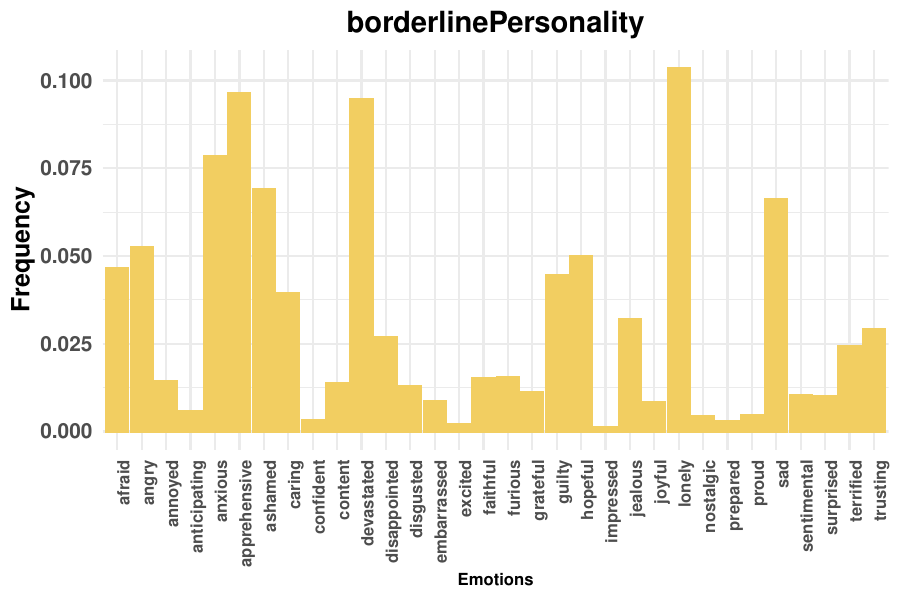}
        \caption{r/bpd (1000 samples)}
        \label{fig:bpd}
    \end{subfigure}
    \hfill
    \begin{subfigure}[c]{0.49\textwidth}
        \centering
        \includegraphics[width=\textwidth]{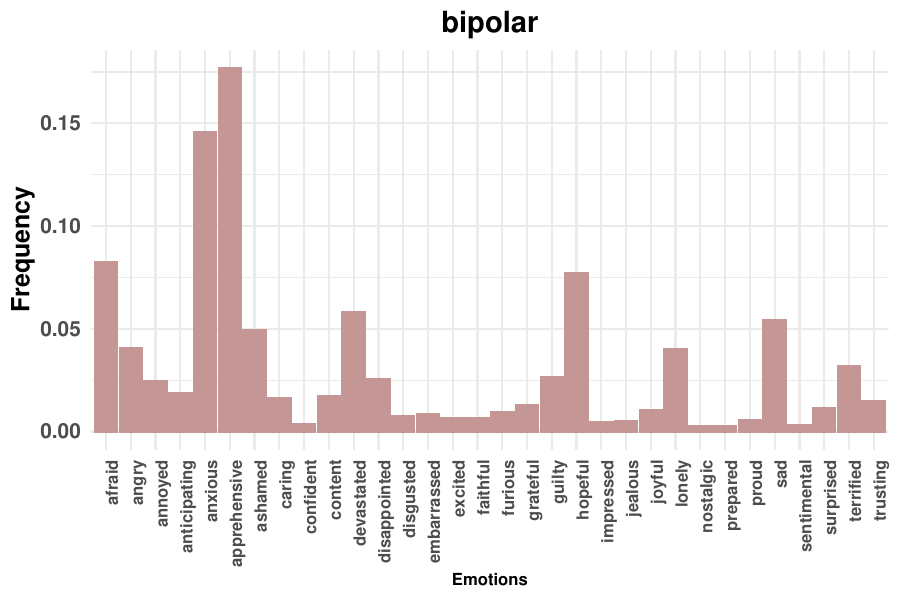}
        \caption{r/bipolar (1000 samples)}
        \label{fig:bipolar}
    \end{subfigure}
    \hfill
    \begin{subfigure}[c]{0.49\textwidth}
        \centering
        \includegraphics[width=\textwidth]{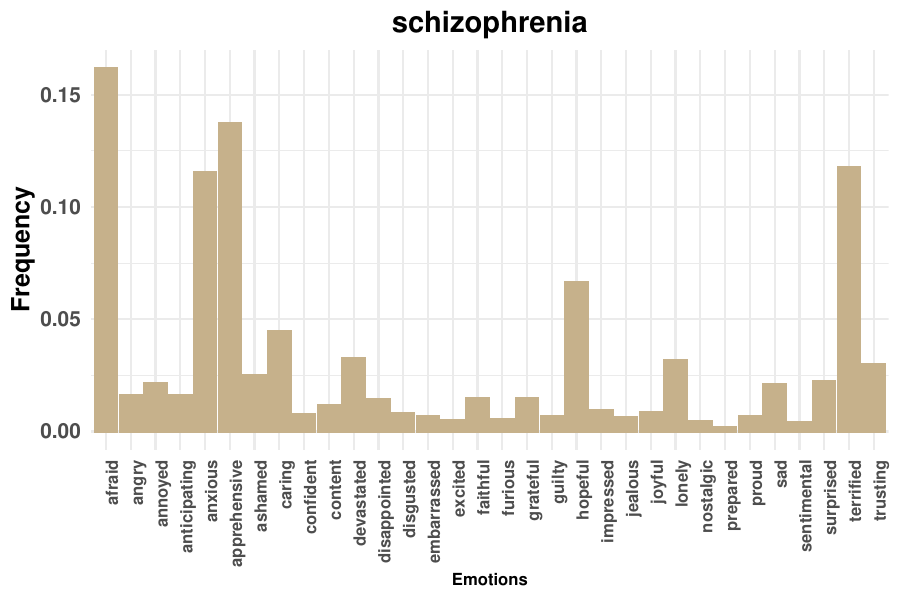}
        \caption{r/schizophrenia (1000 samples)}
        \label{fig:schizophrenia}
    \end{subfigure}
    \hfill
    \begin{subfigure}[c]{0.49\textwidth}
        \centering
        \includegraphics[width=\textwidth]{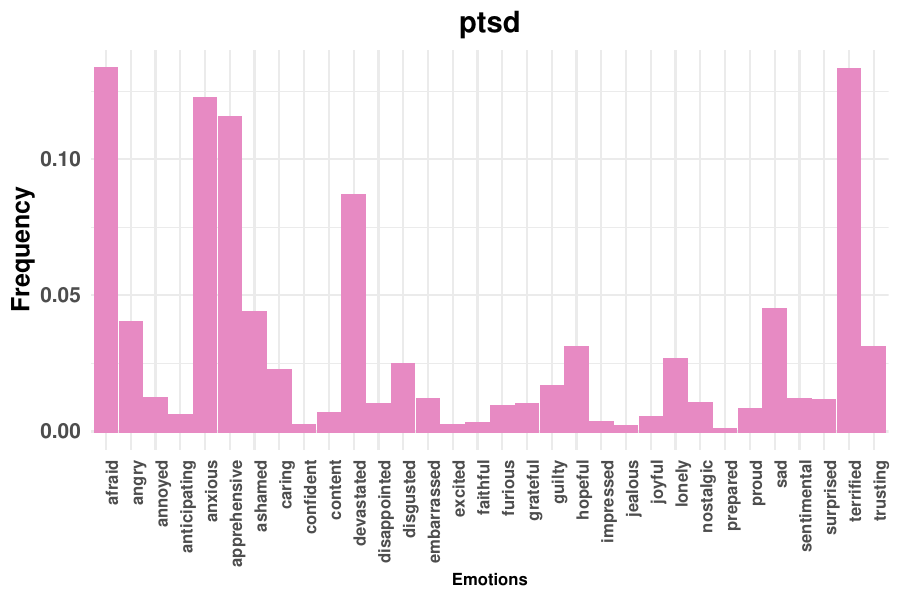}
        \caption{r/ptsd (1000 samples)}
        \label{fig:ptsd}
    \end{subfigure}
    \hfill
    \begin{subfigure}[c]{0.49\textwidth}
        \centering
        \includegraphics[width=\textwidth]{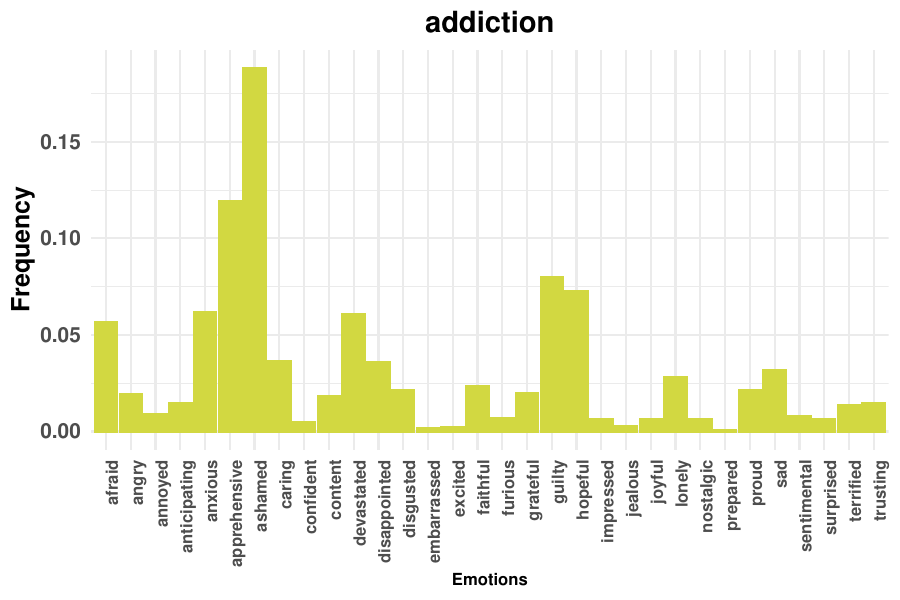}
        \caption{r/addiction (1000 samples)}
        \label{fig:addiction}
    \end{subfigure}
    \hfill
    \begin{subfigure}[c]{0.49\textwidth}
        \centering
        \includegraphics[width=\textwidth]{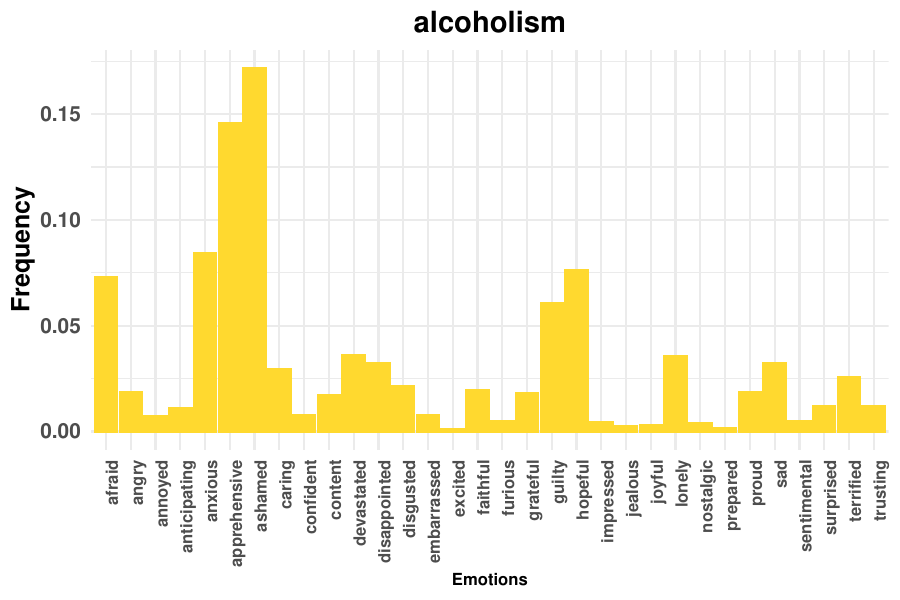}
        \caption{r/alcoholism (1000 samples)}
        \label{fig:alcoholism}
    \end{subfigure}
    \caption{Emotional embeddings of subreddits related to high risk of suicide.}
    \label{distributions}
\end{figure*}

\textbf{r/suicidewatch:} A support forum for individuals experiencing suicidal thoughts, or those concerned about others, offering compassionate responses and sharing feelings. Suicide is closely related to depression, being depression one of the most common disorders among people who die by suicide according to Suicide Risk and Mental Disorders by~\cite{Bradvik2018SuicideRA}.  As expected the embeddings for r/suicidewatch ~\ref{fig:suicide} and r/depression ~\ref{fig:depression} are very similar, the emotional profile is characterized by a disproportionate frequency of extremely negative emotions like `devastated', `sad',  `lonely', and `afraid'.\\

\textbf{r/depression:} A supportive forum for people struggling with depression, where users share their experiences and offer mutual support. According to core symptoms of major depressive disorder by~\cite{Kennedy2008CoreSO}, depression is a common and serious mood disorder that affects a person's feelings, thoughts, and behaviors. It's characterized by persistent feelings of sadness, hopelessness, and a lack of interest or pleasure in activities. Frequent thoughts about death, suicidal ideation, or suicide attempts are also common symptoms in more severe cases.
Figure~\ref{fig:depression} is a visualization of the emotion embedding for depression obtained from the r/depression subreddit. This is indeed characterized by a disproportionate frequency of extremely negative emotions and a lack of positive feelings, the most prominent characteristical emotions are `sad', `lonely', `devastated' and `ashamed'.\\

\textbf{r/bpd:} A subreddit focusing on Borderline Personality Disorder, providing a space for sharing experiences, seeking advice, and finding support. According to Borderline Personality Disorder (BPD): In the Midst of Vulnerability, Chaos, and Awe by~\cite{Kulacaoglu2018BorderlinePD}, Borderline Personality Disorder (BPD) is a complex mental health condition characterized by a pattern of varying moods, self-image, and behavior, marked suicidality and affective instability. These symptoms often result in impulsive actions and problems in relationships with others. Figure \ref{fig:bpd} shows that the emotional embedding is varied across the classified emotional spectrum. Nonetheless the most prominent emotions are `lonely' ,`devastated', `apprehensive', and `anxious'.\\

\textbf{r/addiction:} Focuses on various forms of addiction. And \textbf{r/alcoholism:} A community dedicated to discussing alcoholism. 
According to~\cite{Song2020ComparisonOS} in comparison of Suicide Risk by Mental Illness, addiction, and substance abuse are significant risk factors of suicide. The two embeddings are very similar, Figures \ref{fig:addiction} and \ref{fig:alcoholism} show consistent emotions across the two communities that discuss similar issues, with high frequencies of `ashamed' and `apprehensive'.\\

\textbf{r/schizophrenia:} Dedicated to individuals with schizophrenia, a primary psychotic disorder. People with schizophrenia experience chronic and significant psychotic symptoms, such as hallucinations (seeing or hearing things that are not there) and delusions (false beliefs).\\

\textbf{r/ptsd:} A space for individuals suffering from Post-Traumatic Stress Disorder,  a disorder that usually arises after experiencing or witnessing a traumatic event. Related to psychosis and schizophrenia, according to~\cite{OConghaile2015DistinguishingSF}, in distinguishing schizophrenia from posttraumatic stress disorder with psychosis. In this case, also the embeddings obtained from the two subreddits are similar, with high frequencies of `anxious', `afraid', and `terrified', Figures \ref{fig:schizophrenia} and \ref{fig:ptsd}.\\

\textbf{r/bipolarreddit:} Dedicated to discussions about bipolar disorder. This is primarily a mood disorder characterized by extreme shifts in mood, energy, and activity levels, ranging from manic or hypomanic episodes to depressive episodes. It can have psychotic features, especially during manic or, less commonly, depressive episodes. Figure \ref{fig:bipolar} shows the emotional profile of this disorder, characterized by `apprehensive' and `anxious' feelings.\\

\begin{figure*}
    \centering
    \begin{subfigure}[c]{0.49\textwidth}
        \includegraphics[width=\textwidth]{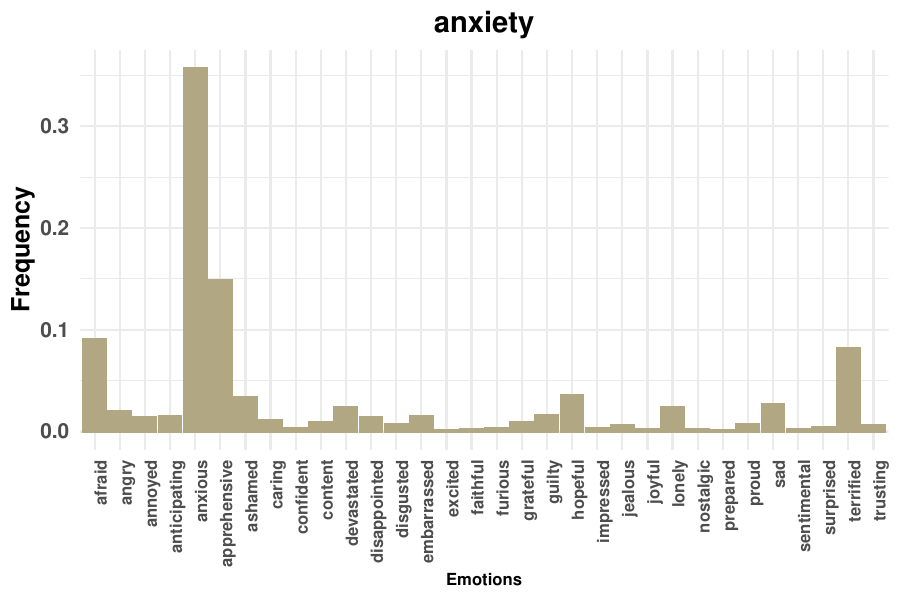}
        \caption{r/anxiety (1000 samples)}
        \label{fig:anxiety}
    \end{subfigure}
    \hfill
    \begin{subfigure}[c]{0.49\textwidth}
        \centering
        \includegraphics[width=\textwidth]{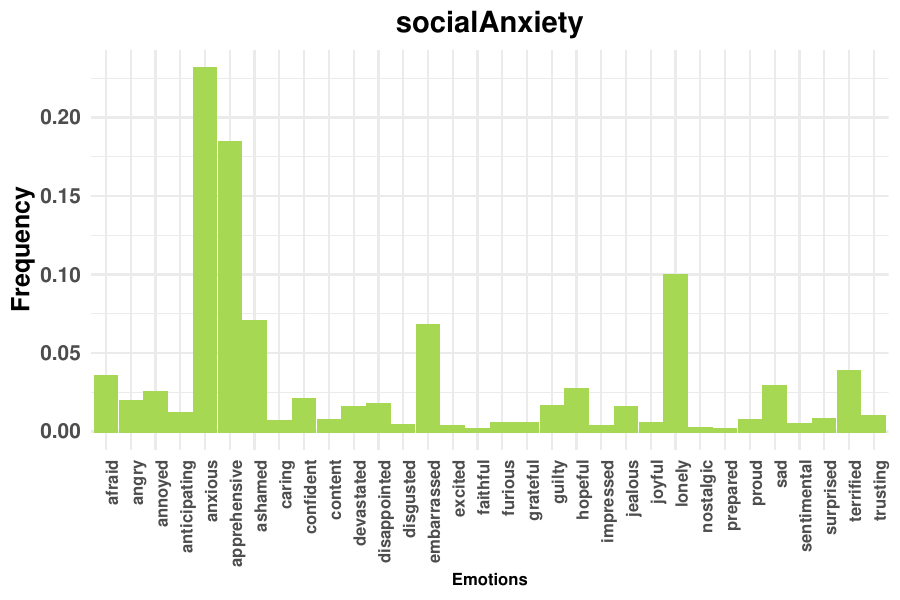}
        \caption{r/socialanxiety (1000 samples)}
        \label{fig:socialAnxiety}  
    \end{subfigure}
    \hfill
    \begin{subfigure}[c]{0.49\textwidth}
        \centering
        \includegraphics[width=\textwidth]{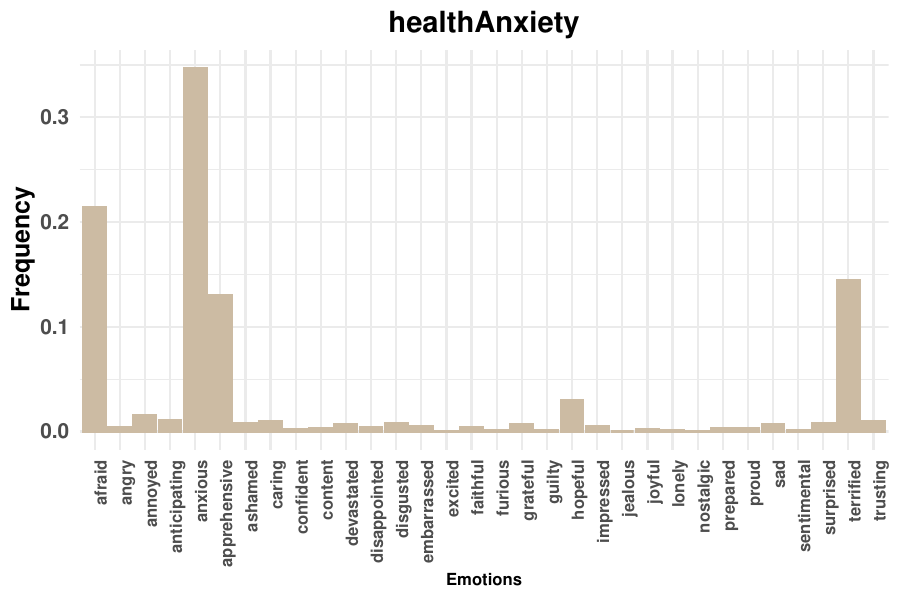}
        \caption{r/healthanxiety (1000 samples)}
        \label{fig:healthanxiety}
    \end{subfigure}
    \hfill
    \begin{subfigure}[c]{0.49\textwidth}
        \centering
        \includegraphics[width=\textwidth]{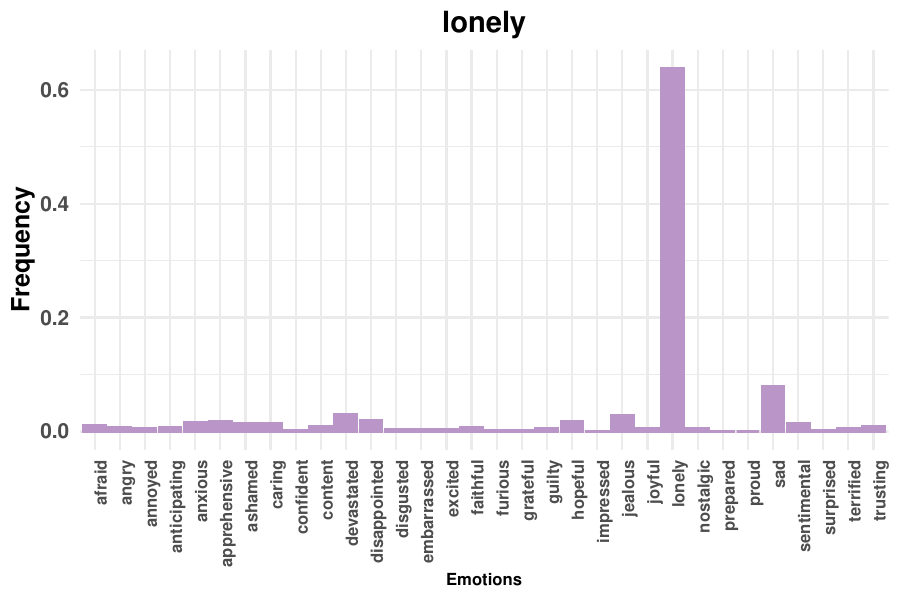}
        \caption{r/lonely (1000 samples)}
        \label{fig:lonely}
    \end{subfigure}
    \hfill
    \begin{subfigure}[c]{0.49\textwidth}
        \centering
        \includegraphics[width=\textwidth]{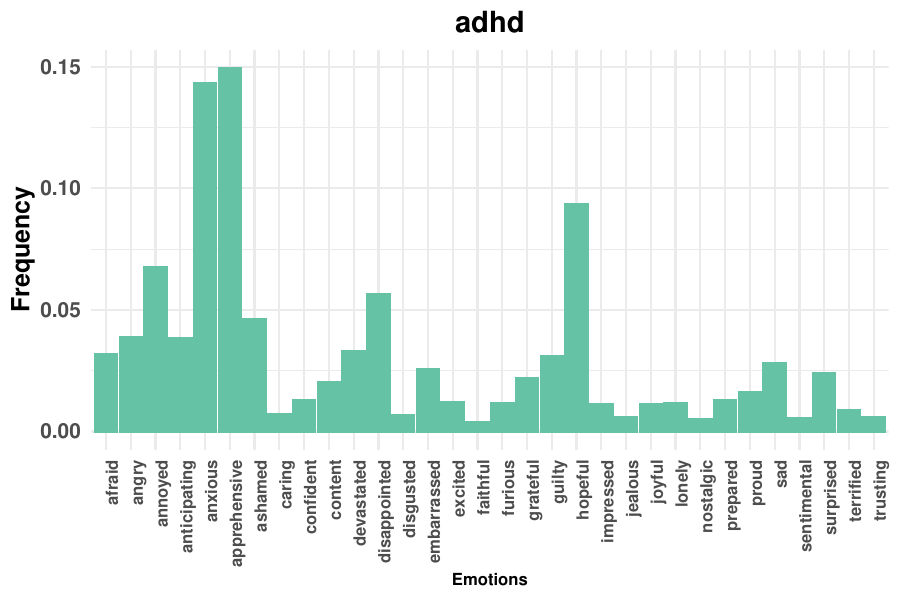}
        \caption{r/adhd (1000 samples)}
        \label{fig:adhd}
        \end{subfigure}
    \hfill
    \begin{subfigure}[c]{0.49\textwidth}
        \centering
        \includegraphics[width=\textwidth]{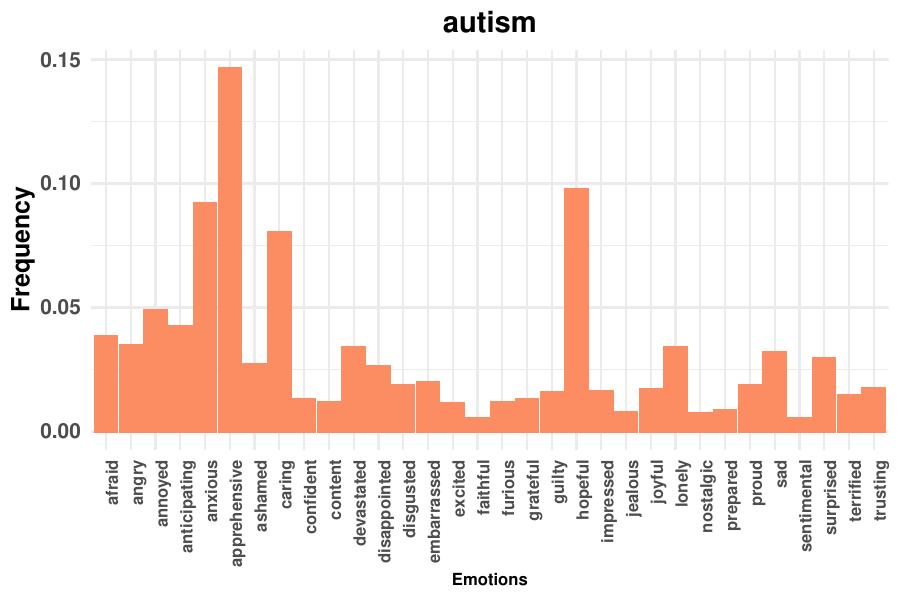}
        \caption{r/autism (1000 samples)}
        \label{fig:autism}
        \end{subfigure}

    \caption{Emotional embeddings of different subreddits.}
    \label{distributions2}
\end{figure*}

\textbf{r/socialanxiety}, \textbf{r/anxiety} and \textbf{r/healthanxiety}: Various subreddits related to anxiety. Indeed Figures \ref{fig:anxiety}, \ref{fig:healthanxiety} and \ref{fig:socialAnxiety} show that the emotions expressed in these communities are dominated by `anxiety'. With the difference that social anxiety also has high frequencies of `lonely' and `apprehensive', and health anxiety `afraid' and `terrified'.\\

\textbf{r/lonely:} A community for those feeling loneliness or isolation. Figure \ref{fig:lonely} show consistent detection of `lonely'.\\

\textbf{r/adhd:} Centered around Attention Deficit Hyperactivity Disorder. Figure \ref{fig:adhd}.\\

\textbf{r/autism:} A community for those affected by autism. \ref{fig:autism}.\\

\newpage

\newpage

\newpage

\end{document}